\documentclass[aps,prb,showpacs,amsmath,amssymb,superscriptaddress,twocolumn]{revtex4-1}
\usepackage[dvipsnames,pdftex,usenames]{color} %options n\'{e}cessaires pour compiler avec pdfLaTex
\usepackage[pdftex]{graphicx}
\usepackage{xspace}
\usepackage{bm}
\usepackage{amssymb}
\usepackage{textcomp,amsmath}

\usepackage[dvipsnames,pdftex,usenames]{color} %options n\'{e}cessaires pour compiler avec pdfLaTex
\usepackage[pdftex,
            bookmarks=true,
            bookmarksnumbered=false,
            bookmarksopen=false,
            colorlinks=true,
            linkcolor=OK-DarkBlue,
            urlcolor=Blue, %Blue or black,
            citecolor=Blue, %Blue or black
            pdfstartview=FitH,
            pdfstartpage=3,
            hyperfootnotes=true,
            pdfhighlight = /N,
            pdfauthor={Olivier Krebs},
            pdfpagemode=None,
            pdftitle={Exchange interaction-driven dynamic nuclear polarization  in  Mn-doped InGaAs/GaAs quantum dots}]
            {hyperref}

\definecolor{OK-DarkBlue}{rgb}{0,0.07,0.4}
\definecolor{OK-DiffNew}{rgb}{0.2,0.04,0.9}

\begin{document}

\preprint{137}


\title{Exchange interaction-driven dynamic nuclear polarization  in  Mn-doped InGaAs/GaAs quantum dots}

\author{O. Krebs}
%\email[]{olivier.krebs@lpn.cnrs.fr}
\affiliation{Centre de Nanosciences et de Nanotechnologies, CNRS, Univ. Paris-Sud, Universit\'{e} Paris-Saclay, C2N – Marcoussis, 91460 Marcoussis, France}
\author{E. Baudin}
\affiliation{Laboratoire Pierre Aigrain, Ecole Normale Sup\'{e}rieure, PSL Research University, CNRS, Universit\'{e} Pierre et Marie Curie, Sorbonne Universit\'{e}s, Universit\'{e} Paris Diderot, Sorbonne Paris-Cit\'{e}, 24 rue Lhomond, 75231 Paris Cedex 05, France}
\author{A. Lema\^itre}

%\thanks{}
\affiliation{Centre de Nanosciences et de Nanotechnologies, CNRS, Univ. Paris-Sud, Universit\'{e} Paris-Saclay, C2N – Marcoussis, 91460 Marcoussis, France}


\date{\today}
\def\xp{$X^{+}$\xspace}
\def\xps{$X^{+^\star}$\xspace}
\def\2xp{$2X^{+}$\xspace}
\def\xm{$X^{-}$\xspace}
\def\x0{$X^{0}$\xspace}
\def\a0{$A^{0}$\xspace}
\def\h{$h$\xspace}
\def\electron{$e^{-}$\xspace}
\def\sp{$\sigma^{+}$\xspace}
\def\sm{$\sigma^{-}$\xspace}
\def\da0{$|\!-\!1\rangle$\xspace}
\def\ua0{$|\!+\!1\rangle$\xspace}
\def\etal{~\textit{et al.}\xspace}
\def\ides{\textit{i.e.}\xspace}

\begin{abstract}
We investigated   optical spin orientation and dynamic nuclear polarization (DNP) in individual self-assembled InGaAs/GaAs quantum dots (QDs) doped by a single Mn atom, a magnetic impurity   providing a neutral acceptor \a0 with an effective spin $J=1$.
%Polarization-resolved photoluminescence spectra of several QDs were recorded under circularly polarized excitation of  positively charged excitons \xp in a longitudinal magnetic field.
We find that the spin of  an electron photo-created in such a quantum dot  can be efficiently oriented by a quasi-resonant circularly-polarized excitation. For the  electron spin levels which are made quasi-degenerate by a magnetic field compensating the exchange interaction $\Delta_e$ with \a0, there is however  a full depolarization due the anisotropic part of the exchange.
%, in which case the  exchange interaction $\Delta_e$ with \a0 induces full depolarization.
%Our analysis reveals that  this effect results from   an anti-crossing $\delta_e$ of the electron spin levels induced by the exchange and the anisotropic environment of \a0.
Still, in most studied QDs, the spin polarized photo-electrons give rise to a pronounced DNP which grows with  a longitudinal magnetic field until a critical field where it abruptly vanishes.  For some QDs, several replica of such DNP sequence are observed at different magnetic fields.
%The extracted Overhauser shift shows a remarkable succession of  increases and abrupt decreases.
This striking behavior is qualitatively discussed as a consequence of different  exchange interactions experienced by the electron, driving the DNP rate via the energy cost of electron-nucleus spin flip-flops.
\end{abstract}

% 78.67.Hc quantum dots
% 75.50.Pp semiconductor magnetic
% 71.70.Gm Exchange interaction, energy level splitting
% 78.55.Cr Luminescence of III-V semiconductors,
% 73.21.La electron states and collective excitations in QDs,
% 76.70.Fz DNP
% 31.30.Gs hyperfine interactions
%78.67.Hc Optical properties of low-dimensional, mesoscopic, and nanoscale materials and structures  : Quantum dots
%78.55.Cr	Photoluminescence, properties and materials : III-V semicondcutors
%71.35.Pq 	Excitons and related phenomena : 	Charged excitons (trions)
%33.80.Be 	Photon interactions with molecules (see also 42.50.-p Quantum optics) : Level crossing and optical pumping
%75.50.Pp 	Magnetic semiconductors


%\keywords{}

\maketitle

\section{Introduction}
Semiconductor quantum dots (QDs) doped by a single or few magnetic impurities have been studied in the last decade in order to investigate the exchange interaction between spin carriers in the quantum regime~\cite{Besombes-PRL04,Leger2005,Leger2006,Leger2007,Kudelski2007,Krebs2009,Trojnar2011,Trojnar2013,Mendes2013}, and the potential of such system as a solid-state quantum bit~\cite{LeGall2009,LeGall2010,LeGall2011,Baudin-PRL11,Goryca2009,Reiter2009,Reiter2011,Besombes2012,Krebs2013,Kobak2014}.  In such QDs the dominant 2-spin interaction is the exchange interaction between  the magnetic dopant and the QD-confined hole (up to a few meV). %, which besides is strongly anisotropic (Ising-like) because of the large spin-orbit coupling in the valence band.
Then comes the electron-hole exchange interaction ($\sim 0.5$ meV) and the exchange interaction between  the magnetic dopant and the QD-confined electron ($\sim0.1$~meV or less). In this context, the hyperfine interaction of the confined carriers  with the $\sim10^4-10^5$ nuclear spins of the QD matrix,  with typical fluctuations in the $\mu$eV  range, turns out to be a small perturbation. However it has been proven to play an  essential role for the spin dynamics  of a single electron in undoped QDs~\cite{Urbaszek2013}, in particular through the ability, under various experimental conditions, to strongly polarize the nuclear spins~\cite{Eble2005,Braun2006,Tartakovskii2007,Maletinsky2007a,Chekhovich2010,Yang2013}. This raises the question whether any such manifestation of the hyperfine interaction between a single electron and the nuclear spin bath can be observed in magnetically doped QDs.

In this paper, we address this issue by focusing more specifically on the electron-\a0 system in single InGaAs/GaAs QDs where \a0 represents the neutral magnetic acceptor, with an effective spin $J=1$, provided by a substitutional Mn impurity in the InGaAs matrix together with its bound  hole\cite{Schneider1987,Marczinowski-PRL07,Kudelski2007}. Our experiments reveal  that  the exchange interaction with the magnetic impurity \a0 drastically perturbs the mechanism of dynamical nuclear polarization (DNP) by a spin-polarized electron. It  leads either to a partial inhibition or, more surprisingly,  to a succession of  DNP increases developing when a longitudinal magnetic field is swept. These observations can be  qualitatively interpreted as a function of the strength and anisotropy  of the electron-\a0 exchange interaction.

\section{Samples and experimental methods}

In the following, we report a set of observations carried out on four distinct Mn-doped InGaAs/GaAs QDs (labelled QD$n$, with $n$=1 to 4) originating from two different samples : QD1 is in a sample consisting   of  a single layer of InGaAs/GaAs QD's with a low p-type residual doping~\cite{Krebs2009}, while QD2, QD3 and QD4 are from a  diode sample where the Mn-doped QD layer is coupled to an electron reservoir enabling the charge control by an applied     electrical bias\cite{Kudelski2007,Baudin-PRL11}.  Let us recall that due to the temperature required for the QD  growth the effective Mn doping  remains quite low with typically less than 1\% of the QDs showing an actual coupling with an \a0 impurity. Thus, the  QDs of the present study were first sought by scanning  some sample areas with a micro-photoluminescence (\textmu-PL) set-up. They were confirmed as Mn-doped QDs thanks to their specific spectral signature   in a magnetic field\cite{Kudelski2007}. Our \textmu-PL set-up relies on a 2-mm~focal length aspheric lens ($0.5$ NA) actuated by piezo-stages and mounted in a split-coil magneto-optical cryostat. The optical excitation is provided either by a HeNe laser or a continuous wave (cw) tunable Ti-sapphire laser. The collected PL is analyzed with a set of linear and quarter-wave plates to resolve its circular $\sigma^+$  or $\sigma^-$ polarization. It is then dispersed by a 0.6-m~focal length double spectrometer equipped with a Nitrogen-cooled CCD array camera providing a  multichannel detection with $\approx$10 meV spectral range and  typical 15~s integration times. All measurements were performed at low temperature ($\leq$5~K) and the magnetic field was applied  parallel to the optical and QD growth axis~$z$.

The QDs have been studied in a regime where they are  positively charged by an additional hole, as evidenced by their magneto-optical signature.~\cite{Krebs2009} In the diode sample, this  relies on an optical charging which takes place when the electron of a photo-created neutral exciton tunnels out of the dot,  due to a high internal electric field.~\cite{Eble2005}  Under optical excitation,   positive trions \xp  (2 holes, 1 electron)  are thus created. In such complex, both holes   are paired in a singlet, so that their spin-related interactions with other particles vanish. The  \xp  spin  thus corresponds  to the spin  of the photo-created electron with eigenstates $|S_{z,e}=\pm1/2\rangle$ (also denoted $\uparrow$ or $\downarrow$), which interacts  with the \a0 spin and the QD nuclear spins during the trion lifetime. Thanks to the optical selection rules of  trions in QDs (inherited from the heavy hole spin-orbit coupling), it can be analyzed via the PL circular polarization which reads $\mathcal{P}_\text{c}=2\langle S_{z,e}\rangle=(I_{\sigma-}-I_{\sigma+})/(I_{\sigma-}+I_{\sigma+})$ where $I_{\sigma\pm}$  is the PL intensity detected in $\sigma\pm$ polarization.

\section{Spin polarization in magnetic field}

In undoped QDs, exciting \xp trion with circularly polarized light offers a direct mean to investigate the hyperfine interaction with the nuclei and possibly to induce efficient dynamical nuclear polarization \cite{Urbaszek2013}. For example, changes in PL circular polarization  can  reveal electron spin relaxation induced by the transverse fluctuations of the so-called Overhauser field (\ides the nuclear spin polarization), whereas the spectral splitting of the PL circular components in zero field, or its shift with respect to normal Zeeman splitting in a magnetic field $B_z$, reflects the average value of this Overhauser field along $z$. To apply the same approach  to the case of Mn-doped QDs, we first need  to carefully analyze the electron-\a0 system, namely its level structure, the spin eigenstates and the intrinsic spin polarization which may develop due to magnetic field and exchange interactions with \a0 in the absence of optically induced spin orientation or pumping.

\begin{figure}[ht]
\includegraphics[width=0.48\textwidth]{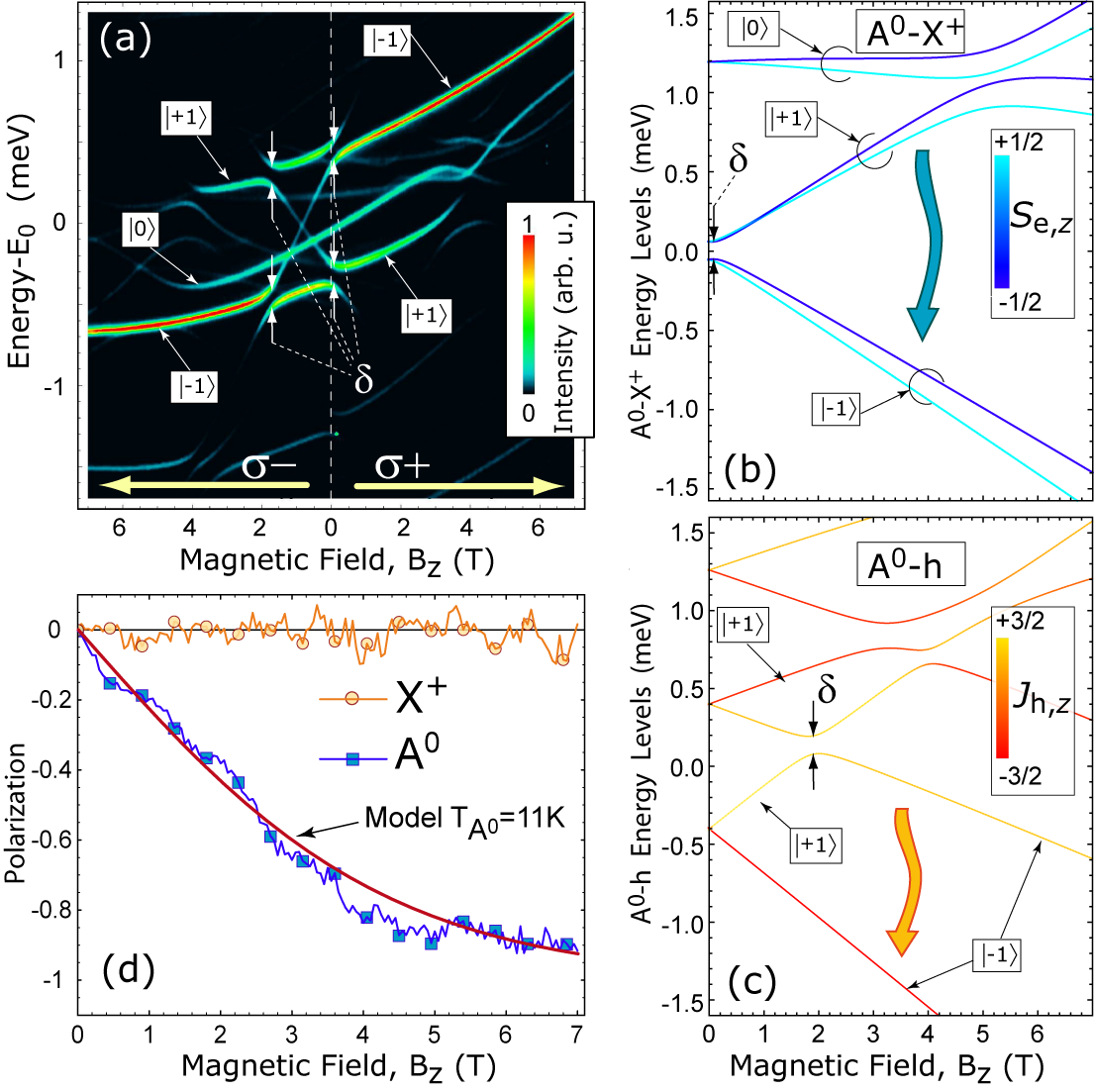}
\caption{\label{Fig1} (a)  QD1 PL spectra measured  in circular polarization ($\sigma^+$ or $\sigma^-$ as indicated) as a function of the  magnetic field $B_z$  under non-resonant unpolarized excitation. $E_0=1.341$~eV and T=5K.  (b),(c) Calculated energy levels of the initial state (A$^0$-X$^+$) and final state (A$^0$-h) providing a good simulation of (a) (not shown). The wavy arrows illustrate the expected relaxation  leading to a large spin polarization. (d) Polarization of A$^0$ spin  and \xp PL  against magnetic field. The solid line is calculated according  to Brillouin's model of thermalization for a 1/2 spin. }
\end{figure}
Figure~\ref{Fig1} presents such preliminary investigation of QD1 achieved under non-polarized and non-resonant excitation (633 nm HeNe laser line, 10~\textmu W incident power).  In Fig.~\ref{Fig1}(a), the characteristic  polarization-resolved magneto-PL spectra of such Mn-doped InGaAs QD is shown : it consists of two main lines, corresponding to the trion transitions with constant  \a0  spin states $|\!\pm1\rangle$,  which anticross with two weaker lines corresponding to ``forbidden'' transitions where the \a0 spin is flipped ($|\!\pm1\rangle \rightarrow |\!\mp1\rangle$). Altogether they form a  remarkable X-pattern. The  theoretical levels corresponding to these transitions are plotted in Fig.~\ref{Fig1}(b),(c).  Detailed discussions about such images and the model Hamiltonians enabling us to calculate the levels can be found in Ref.~\onlinecite{Krebs2009,Krebs2013}. Here we mostly focus on the \a0 and trion spin polarization which results from the thermal relaxation taking place both  in the \a0-\xp configuration  (transition initial state) and  in the \a0-hole configuration (final state), as illustrated by wavy arrows in Fig.~\ref{Fig1}(b),(c). The experimental trion polarization (\ides electron spin polarization) can be easily deduced from Fig.~\ref{Fig1}(a) by integrating separately the $\sigma^+$ and $\sigma^-$  PL intensity over a typical 3~meV spectral range. To estimate the \a0 spin polarization, namely the ratio $(p_{+1}-p_{-1})/(p_{+1}+p_{-1})$ where $p_{\pm1}$ represents the \a0 population in the state $|\pm1\rangle$, we extract with an appropriate line fit  the total intensity of the PL lines  associated either to a  $|\!+1\rangle$  or a $|\!-1\rangle$ state, which are assumed to be proportional to the corresponding populations.
%assumed supposed to be proportional to the corresponding populations.
%we  have to carefully select the lines corresponding respectively to the  $|\!+1\rangle$  and $|\!-1\rangle$ states  as their intensities is assumed to be proportional to the corresponding population. for both $\sigma^+$ and $\sigma^-$ polarizations and integrate their intensities.
The results are shown in Fig.~\ref{Fig1}(d).

As previously observed,\cite{Kudelski2007,Besombes-PRL04} the \a0 spin acquires a strong polarization (negative in positive fields) when the  magnetic field $B_z$ increases. This behavior can be fairly well reproduced by the Brillouin function of a 1/2 spin,  $B_{1/2}(B_z)=\tanh(g_{A^0} \mu_\text{B} B_z/k_\text{B}T_{A^0})$, where $g_{A^0}=3.6$ is the \a0 g-factor determined from the X-pattern in Fig.~\ref{Fig1}(a), $\mu_\text{B}$ is the Bohr magneton, $k_\text{B}$ is the Boltzman constant and $T_{A^0}=11$~K is the \a0 temperature. This  temperature, considered here as a fitting parameter, is slightly higher than the lattice temperature (5~K), likely due to the optical excitation and recombination with \a0 spin-flip. It still indicates an efficient spin relaxation to the two lowest levels  of \a0-\xp leading to an \a0 spin polarization of -90\% at $B_z=7$~T.

In contrast, the \xp polarization remains essentially equal to zero, although the electron thermal polarization in \xp should amount to 25\% at 7~T and 5~K, according to the electron $g$-factor $g_e=-0.48$ in QD1 [deduced from Fig.~\ref{Fig1}(a)]. This indicates that the electron spin relaxation is likely inefficient during the \xp lifetime. This absence of \xp polarization also implies that the polarization of the resident hole, which is presumably quite  high because of the large \a0-hole exchange interaction [see Fig.~\ref{Fig1}(c)], is not transferred to the electron when an \xp is created non-resonantly.

From these preliminary observations, it can thus be assumed that the \xp polarization will reflect with fidelity  the spin state of the electron captured  or photo-created in the QD and its intrinsic subsequent evolution, in particular under the influence of the hyperfine interaction or the electron-\a0 exchange. In that respect, the   field $|B_\delta|\approx79$~mT  of the $\delta$ anticrossing in Fig.~\ref{Fig1}(b) provides a direct  estimate of the exchange strength $\Delta_{e-A^0}= 2 g_{A^0} \mu_\text{B} |B_\delta|\approx33$~\textmu eV. In principle, this should protect the electron spin from  the nuclear spin bath fluctuations. However, near zero field, the actual electron spin splitting  is significantly reduced because of the  strong mixing of the $|\!+1\rangle$ and $|\!-1\rangle$ \a0 states due to the  $\delta/2\approx 72$~\textmu eV coupling.  It is thus difficult to predict if the small hyperfine interaction with the nuclei will induce some spin relaxation and/or initiate a nuclear spin polarization which then could be amplified by  a magnetic field  as observed for undoped QDs\cite{Urbaszek2013}.

\section{Correlated \a0-trion spin polarization}

\begin{figure}[ht]
\includegraphics[width=0.48\textwidth]{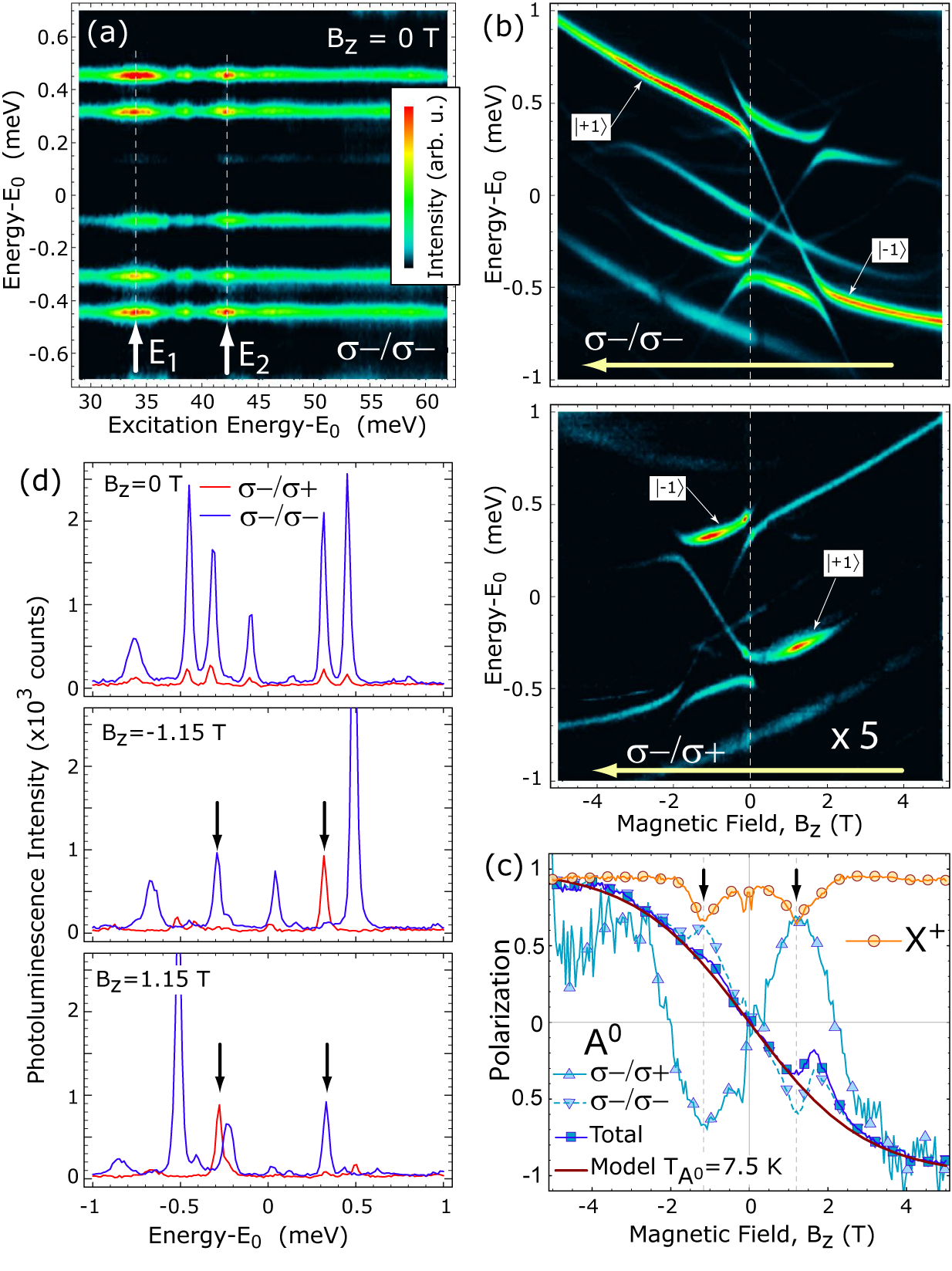}
\caption{\label{Fig2} (a) QD1  PL spectra measured in co-polarized configuration ($\sigma-/\sigma-$) as a function of the excitation energy. (b) PL spectra under quasi-resonant  $\sigma-$ excitation  [resonance E$_1$ in (a)]   as a function of a decreasing  magnetic field $B_z$   and measured in $\sigma-$ (top) or $\sigma+$ (bottom, $\times$5-amplified) polarization. (c) Polarization of A$^0$ spin for $\sigma+$ ($\vartriangle$), $\sigma-$ ($\triangledown$) or both measurements ($\square$) and  \xp PL ($\bigcirc$)  against  magnetic field deduced from (b). A$^0$ polarization conditioned to the co- or cross-polarized measurements shows strong correlations with the trion PL polarization. (d) PL spectra at three specific magnetic fields. The  co- and cross-polarized lines marked by arrows exhibit the same intensity at $B_z=\pm$1.15~T.
}
\end{figure}
In order to observe a high degree of \xp circular polarization, InGaAs QDs have to be excited more resonantly, at least below the $\sim$1.42~eV wetting layer bandgap.  We first performed the PL excitation spectroscopy of QD1 with a tunable cw Ti:Sapphire in circular polarization, see Fig~\ref{Fig2}(a). Note that the PL intensity has been normalized by the incident laser power which was  varying  from $\sim$1~mW to $\sim$4~mW when increasing the excitation energy. Two resonances $E_1$ and $E_2$ were found respectively 34~meV and 42~meV above the central  PL energy $E_0=1.341$~eV of QD1, both providing a noticeable polarization above 80\% in zero field.  This first result indicates that, like in undoped QDs, the electron spin of \xp is well protected from the relaxation caused by the fluctuations of the hyperfine interactions, whatever is the reason (strong electron-\a0 exchange or  DNP-induced Overhauser field). Note that the above resonances are relatively large ($\geq 1$~meV) and do not select any specific \a0 spin state, at least in zero field.

We then investigated how the \xp optical orientation evolves in a magnetic field. Figure~\ref{Fig2}(b) shows the magneto-PL images obtained under $\sigma^-$ excitation at $E_1$ when varying the magnetic field from +5~T to -5~T for both co-polarized ($\sigma^-/\sigma^-$) and cross-polarized ($\sigma^-/\sigma^+$) detection. The latter, which is plotted  with a x5-amplified color-scale, evidences  noticeable enhancements around the two magnetic fields of $\pm 1.15$~T. This likely indicates local reductions of  \xp polarization around these fields, which is confirmed by extracting from the integrated spectra the whole \xp polarization. Two pronounced dips of about 20\% amplitude appear around these fields as indicated by arrows in Fig.~\ref{Fig2}(c).

What is quite remarkable is that the enhancement of the $\sigma^-/\sigma^+$ PL signal is clearly  correlated to a specific $|\!+1\rangle$ or $|\!-1\rangle$ \a0 state. This behavior looks like a strong polarization of \a0 opposite in sign to that resulting from the usual thermalization. To analyse these observations more quantitatively, we extracted  in Fig.~\ref{Fig2}(c) the \a0 polarization   separately for the cross- or co-polarized configurations, as well as the total \a0 polarization  deduced from both sets of measurements. They are plotted together with a Brillouin function   at an effective temperature $T_{A^0}=7.5$~K. Note that due to the overall strong PL polarization, the total \a0 polarization is very similar to the \a0 polarization extracted from the co-polarized configuration.   The strong inversion of \a0 polarization in cross-polarized configuration, up to $\sim\pm70$\% at $\pm1.15$~T, appears to be essentially compensated by a small increase of the normal polarization (in absolute value) in the co-polarized one: the total \a0 polarization indeed no longer shows any significant anomaly with respect to the Brillouin function at these fields. There is still  a polarization reduction  around $+1.7$~T, but  this effect most likely results  from the $\delta$ anticrossing which induces the  total mixing of the  $|\!+1\rangle$ and $|\!-1\rangle$ \a0 states occurring  for the $\sigma^-$ polarized transition toward  the  hole spin state $J_z=+3/2$, see Fig.~\ref{Fig1}(c). Incidentally, this absence of anomaly on the total \a0 polarization associated to the depolarization of \xp  allows us to exclude  an electron-induced spin orientation of \a0, in contrast to the Mn spin in magnetic  CdTe/ZnTe QDs~\cite{LeGall2009}, as the cause of this correlation.
%The enhancement of  \xp spin relaxation at $\pm1.15$~T simply takes place  for  a given $|\!\pm1\rangle$ \a0 state respectively.

To elucidate the origin this intriguing correlation,  the calculated \a0-\xp levels in Fig.~\ref{Fig1}(b) turn out very helpful. It is noteworthy that  the two electron spin levels (with $S_{e,z}=\pm1/2$) associated to the $|\!+1\rangle$ \a0 state remain very close up to $\sim$1.5~T with most probably a  crossing at about  1~T, corresponding to the exchange field  $B_\Delta=\Delta_{e-A^0}/g_e \mu_\text{B}$. The symmetrical situation (not shown) holds at about -1~T  for the levels associated to the $|\!-1\rangle$ \a0 state. Since  the spin  relaxation by energy conserving mechanisms (like hyperfine interaction) is favored when the electron spin splitting vanishes, such crossings certainly point to the origin of our observations.  Experimentally, the electron spin splitting can not be directly  observed because the two allowed transitions from these $S_{e,z}=\pm1/2$ levels  with respectively a $J_z=\mp3/2$ hole are split by the strong \a0-hole exchange in the final state. We can still compare the intensities of the corresponding lines to assess  more quantitatively the amount of electron spin relaxation between the two levels. Figure~\ref{Fig2}(d) shows three cross-sections of the magneto-PL images on a common vertical scale. Whereas in zero field the  \xp polarization is  strong for all the QD1 transitions,  at  $\pm1.15$~T the two \xp lines corresponding to the   $|\!\pm1\rangle$ \a0 state (marked by a dark arrow) exhibit  essentially the same intensity in cross- and co-polarized measurements. Obviously, the electron spin relaxation taking place at these crossing points is very high (if not total).

This result  actually discards the hyperfine interaction with the nuclei as  the dominant mechanism for the spin relaxation, because the depolarization by nuclei, even for degenerate electron spin states,  would be  limited to $\sim50$\% of its initial value over  the $\sim1$~ns trion lifetime\cite{Urbaszek2013}. By inspecting closer the two calculated levels associated to a same $|\!\pm1\rangle$ \a0 state, we found out that actually they  anticross  by an energy $\delta_e\approx9$~\textmu eV for QD1. In this region,  the \xp eigenstates  are totally mixed  spin states, and since $\delta_e$ is significantly larger than the \xp natural width, the average spin polarization vanishes. The origin of this anticrossing is obviously related to the electron-\a0 exchange and to the lack of perfect rotational symmetry of the system, in a way similar to the previously observed dark-bright mixing of the \a0-hole spin levels\cite{Kudelski2007,Krebs2009,Goryca2010}. This will be discussed in more detail below with  the measurements on QD2 for which the $\delta_e$ anticrossing could be experimentally resolved.% (phrase \`{a} reprendre en fonction de la suite).

\section{ Fine anticrossing of \a0-trion levels}

\begin{figure}[ht]
\includegraphics[width=0.48\textwidth]{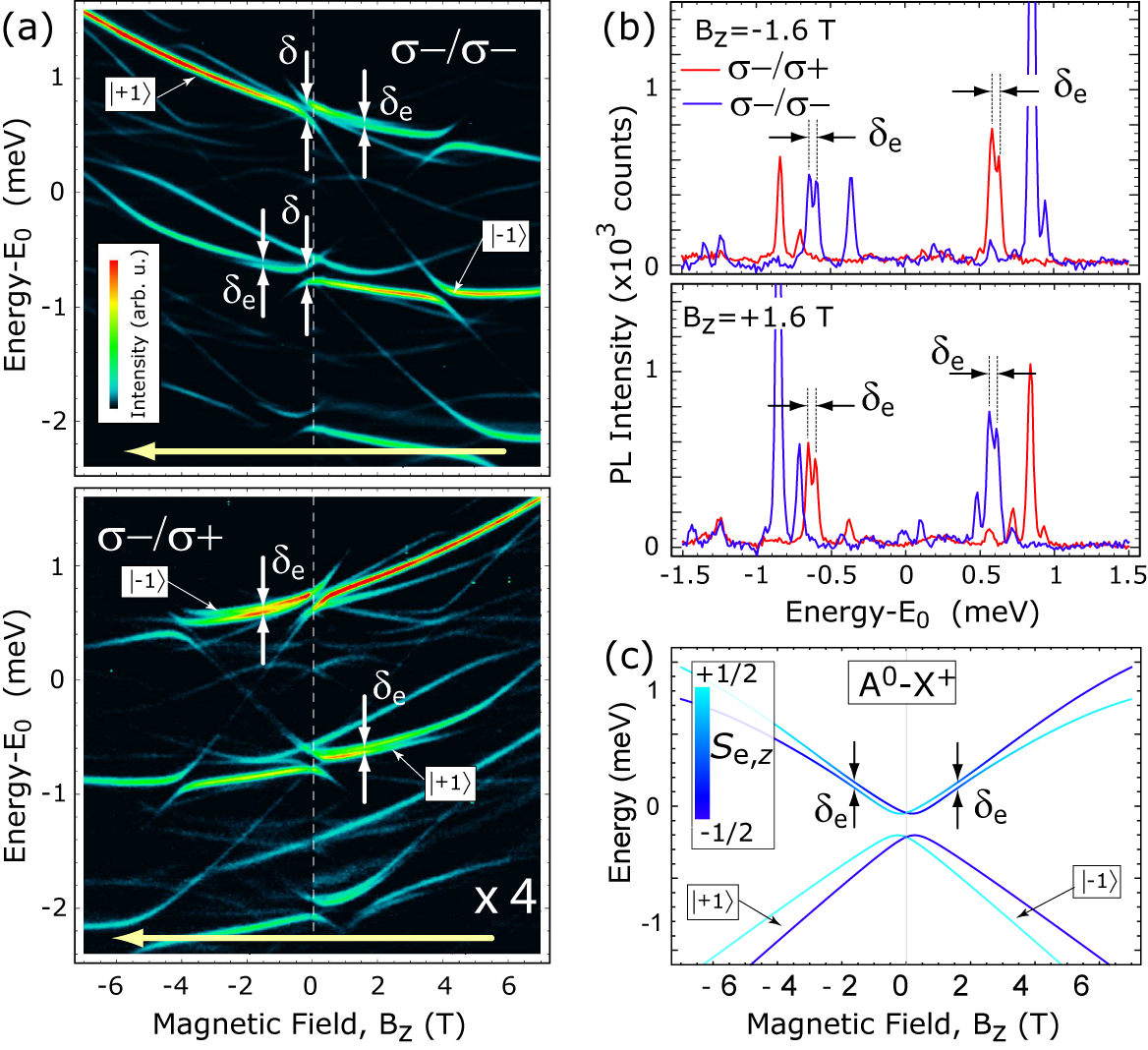}
\caption{\label{Fig3} (a) QD2 PL spectra under quasi-resonant  $\sigma-$ excitation at $E_0+60$~meV  as a function of a decreasing  magnetic field $B_z$   and measured in $\sigma-$ (top) or $\sigma+$ (bottom, $\times$4-amplified) polarization. ($E_0=1.281$~eV). White arrows point out spectrally resolved anticrossings. (b) PL spectra at $B_z=\pm$1.6~T showing the  splitting $\delta_e\approx47$~\textmu eV of the anticrossing transitions responsible for the reduction of PL polarization. (c) Calculated energy levels of the corresponding e-A$^0$ states.
}
\end{figure}

QD2 in Fig.~\ref{Fig3}(a) exhibits a particularly strong exchange interaction with the \a0 impurity characterized by a $\simeq$1.4~meV splitting between the $|\!+1\rangle$ and $|\!-1\rangle$  lines. From the field $|B_\delta|=215$~mT  of the $\delta$-anticrossing and the \a0  g-factor  $g_{A^0}=2.8$ deduced from the X-pattern, we can extract the contribution due to the  electron-\a0 exchange energy to $\Delta_{e-A^0}= 2 g_{A^0} \mu_\text{B} |B_\delta|\approx70$~\textmu eV.  Although it is about twice larger than for  QD1, there is no strong enhancement of the cross-polarized PL lines associated to the $|\!+1\rangle$ or $|\!-1\rangle$  \a0 states near the field where the levels are expected to anti-cross. Instead, we observe a clear splitting of the $|\pm1\rangle$ lines,  both in co- and cross-polarized configurations, over the magnetic field range from basically 0~T up to about $\pm3$~T. The PL spectra  in Fig.~\ref{Fig3}(b) illustrate this feature at the fields of  $\pm$1.6~T where the split lines of  each $\delta_e$-doublets have roughly the same intensity and  the $\delta_e$ splitting reaches a minimum value of 47~\textmu eV.

These observations are in good agreement with the prediction of our simplest spin model\cite{Krebs2013} based on an effective J=1 for \a0, as shown by the calculated levels in Fig.~\ref{Fig3}(c). An important parameter of this model, which notably explains the drastic difference between QD1 and QD2, is the inclination by an angle $\theta_s$  of the dominant strain field experienced by the \a0 impurity with respect to the QD growth axis. For QD2 the angle $\theta_s$ deduced from the simulation of the magneto-PL images is estimated to be 33$^\circ$ while it amounts to only  13$^\circ$ for QD1. It allows for the coupling of the $|\!+1,\uparrow\rangle$ ($|\!-1,\downarrow\rangle$) \a0-electron state to   $|0,\uparrow\rangle$ ($|0,\downarrow\rangle$) which is also coupled by the (Heisenberg-like) electron-\a0  exchange to  $|\!+1,\downarrow\rangle$ ($|\!-1,\uparrow\rangle$). Both terms induce a coupling of the $\uparrow$ and $\downarrow$  electron spin states associated to the same $|\!+1\rangle$ or $|\!-1\rangle$ state, which can be seen as an effective in-plane magnetic field $\delta_e/(g_e \mu_\text{B})$ and gives rise to the observed anti-crossing. To the first order in perturbation theory, we indeed find that the $\delta_e$ splitting is given by $2 \Delta_{e-A^0} \sin 2 \theta_s/(3+\cos 2\theta_s)$ which provides good estimates of the   $\delta_e$ splitting for QD1 (7.4~\textmu eV) and QD2 (40~\textmu eV) from the fitted  angle $\theta_s$.

%donner les valeurs de ces \'{e}changes d'apr\`{e}s les \delta sur la figure, comparer peut\^{e}tre avec les 2-6 et dire que \c{c}a soul\`{e}ve peut\^{e}tre une question mais que c'est surement l'interaction au trou du a0 qui domine. d\'{e}crire succinctement fig3a, par rapport \`{a} la polar moyenne aux effets de d\'{e}polar moins visible mais  cause du d\'{e}doublement clairement visible. commenter le rapport B_\delta/B_A0 vs g_e/g_A0, pour convaincre le lecteur que tout se tient. Enchainer sur les cross sections qui permettent de r\'{e}soudre \delta_e, \'{e}norme 48µeV!. Commenter l'in\'{e}galit\'{e} visible des intensit\'{e}s : puret\'{e} de A0 dans l'\'{e}tat final n'est pas le m\^{e}me pour les deux doublets. (remarque on ne peut pas passer continument de +1/2 \`{a} -1/2 sans passer par 0, \`{a} l'ac ou pas loin le spin vaut forc\'{e}ment 0) la figure 3c illustre bien l'anticroisement qui annulle le spin de e ! le mod\`{e}le confirme bien les observations (ordre de grandeur \`{a} v\'{e}rifier). essayer de trouver une formule simple ou sinon v\'{e}rifier quels sont les termes pr\'{e}dominants.
%conclure sur le fait que le fort \'{e}change e-a0 semble inhiber fortement la dnp, soit par annulation du spin, soit par \'{e}cartement des niveaux. Dire que c'est ce qui se passe pour QD1 et QD2 mais pas totalement en champ nul, donner les OHS mesur\'{e}s, mais enchainer sur Fig 4...

From the above analysis,  it seems  obvious that the electron-\a0 exchange in Mn-doped QDs must strongly inhibit the DNP mechanism   with  spin-oriented electrons:   to experience a spin flip-flop with a nucleus with a reasonable probability the two electron spin states have to be close enough in energy \cite{Urbaszek2013}, but in this case the electron spin, optically oriented along $z$, tends to vanish due to the  $\delta_e$-induced coupling, see Fig.~\ref{Fig3}(c). In zero field, this conclusion can be slightly revised because  of the strong mixing of the $|\!+1\rangle$ and $|\!-1\rangle$ states. The electron-\a0 eigenstates form two Kramers doublets split by $(\delta^2+\Delta_{e-A^0}^2)^{1/2} $ and reading  $|+,\uparrow \rm{or} \downarrow\rangle$ and $|-,\uparrow \rm{or} \downarrow\rangle$ where $|\pm\rangle\simeq(|\!+1\rangle\pm|\!-1\rangle)/\sqrt{2}$. Within each of these doublets the electron can experience spin flip-flops with a nucleus  without any energy cost, leading possibly  to DNP. In the above experiments, by comparing the co- and cross-polarized spectra in  zero field, we could indeed observe for both QD1 and QD2 a finite Overhauser shift amounting to about 15~\textmu eV,  very similar to that observed in undoped QDs. Furthermore, after a careful analysis of the data as detailed below, it turns out that this shift actually survives up to   $\approx$0.5~T while increasing up to $\approx$25~\textmu eV, and then abruptly  vanishes, most likely because of the $\delta_e$-induced coupling.

In undoped QDs, it is usual to observe an increase of  the Overhauser shift, growing approximately like the electron Zeeman splitting $|g_e \mu_\text{B} B_z|$, up to fields above 4~T where it  can  reach more than 100~\textmu eV. This growth is interrupted when the depolarization mechanisms get more efficient than the maximum DNP rate obtained for a zero electron-spin splitting (namely when the Overhauser shift gets exactly  compensated by the Zeeman effect). This results in an abrupt collapse of the nuclear field  which then gives  rise to  spectral jumps  of the $\sigma^+$ or  $\sigma^-$ PL lines making a clear fingerprint of the DNP vanishing  in the magneto-PL images. In Fig's.~\ref{Fig2}(b)~and~\ref{Fig3}(a), these spectral jumps are hardly visible. However, for QD1, by changing slightly the excitation conditions we could create a DNP regime working up to $\approx$2~T as discussed in the following. For QD2 the splitting $\delta_e$ was definitely too large to enable a large Overhauser field to develop, at least in the ranges of temperature, excitation energy and power  we have explored.

\section{Dynamic nuclear polarization in Mn-doped QDs}

\begin{figure}[ht]
\includegraphics[width=0.48\textwidth]{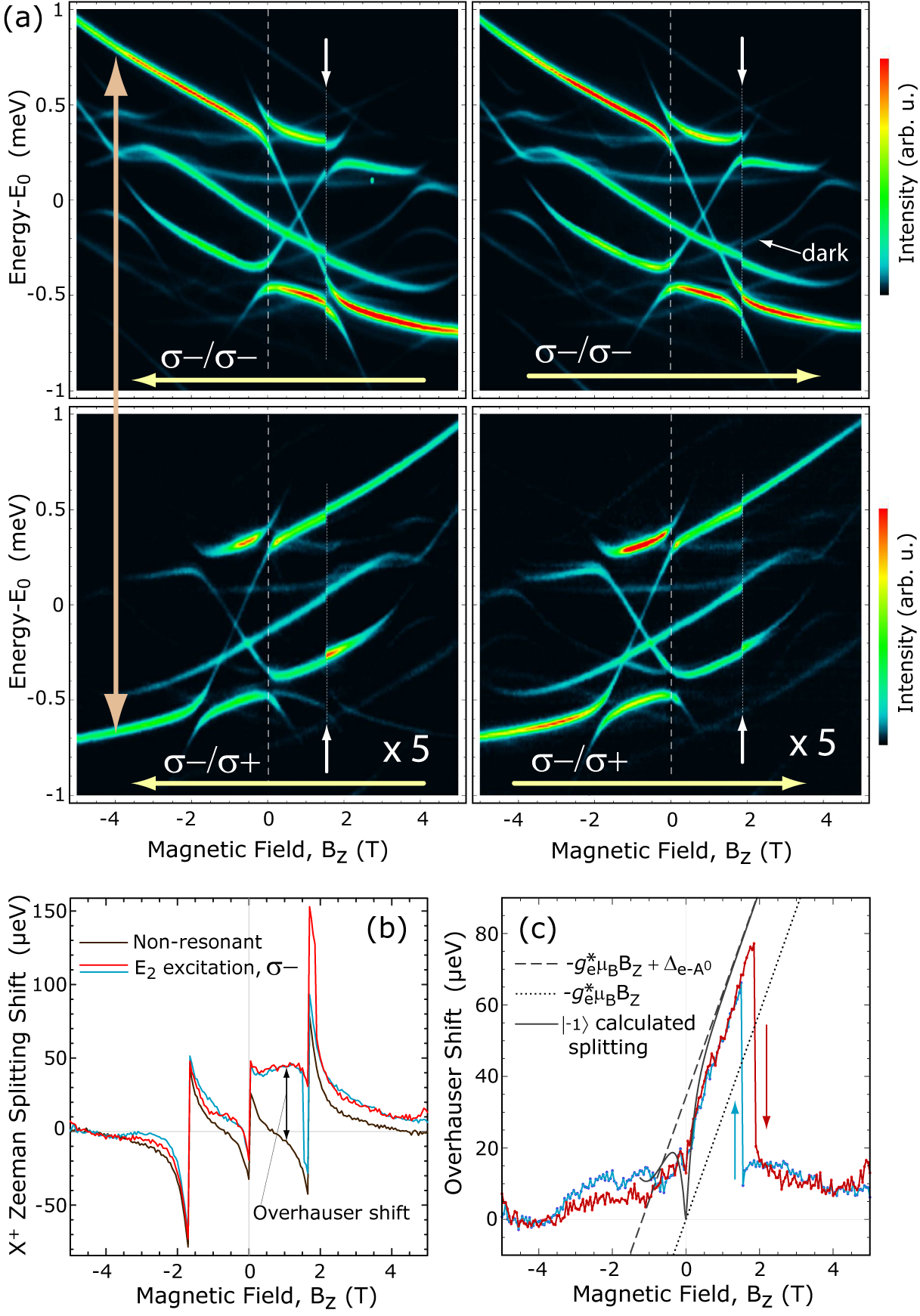}
\caption{\label{Fig4} (a) QD1 PL spectra under quasi-resonant  $\sigma-$ excitation  [resonance E$_2$ in Fig.\ref{Fig2} (a)]   as a function of a decreasing (left) or increasing (right) magnetic field $B_z$   and measured in $\sigma-$ (top) or $\sigma+$ (bottom) polarization. (b) Shifts of  the linear Zeeman splitting deduced from the dominant $\sigma+$ and $\sigma-$ lines of each spectrum under non-resonant excitation or quasi-resonant  $\sigma-$ excitation. (c) Overhauser shift  deduced from (b) for both field sweep directions. % The electron spin splitting is indicated  for the calculated $|\!-1\rangle$ \a0 states (solid line),  and assuming no exchange ($-g_e\star \mu_\text{B} B_z$, dotted line).
}
\end{figure}

Figure \ref{Fig4}(a)  shows the fingerprints of a large Overhauser field which  builds up  in QD1 when the excitation energy is set to the second resonance E$_2$ of Fig.~\ref{Fig2}(a).  The white arrows  indicate noticeable jumps of the $\sigma^+$ ($\sigma^-$) PL lines towards higher (lower) energies for an increasing magnetic field, or towards  lower (higher) energies in a decreasing magnetic field. Let us recall that due to the relative long integration times of each spectrum (10~s) the magnetic field was varied step by step with 50~mT increments, during which the $\sigma^+$ and $\sigma^-$ spectra were successively measured. It is therefore normal to observe the jumps at the same magnetic field for a given field sweeping. Conversely, the jumps take place at different magnetic fields, respectively  at 1.9~T (1.5~T) for an increasing (decreasing) field, revealing the  non-Markovian character of the DNP process  like observed in undoped QDs\cite{Urbaszek2013}. The reason for the drastic change in DNP regime for the two investigated resonances is not clear. Still, we suspect it might be related to the high sensitivity of the DNP mechanism to the effective broadening of the electron spin levels\cite{Urbaszek2013}, which would be  increased when exciting at higher energy.

For  Mn-doped QDs, the Overhauser shift is  difficult to  extract precisely,   because of the numerous spectral lines  which experience  several crossings or anti-crossings as a function of the field. Our method to solve this issue consists in determining by a local Gaussian fit the energy  of only  the $|\!+1\rangle$  ($|\!-1\rangle$) line in negative (positive) fields. In case of ambiguity because of an anticrossing (e.g. near zero field), we retain only the most intense line in order to keep  a single line for each spectrum. The energy difference between the same $|\pm1\rangle$ lines measured in co- and cross-polarized configurations provides, after subtracting a constant exchange energy,  the  \xp Zeeman splitting together with  the Overhauser shift, but including also noticeable spectral deviations due  to the $|\pm1\rangle$ anti-crossings near zero field or  $\pm$1.8~T. A reference measurement is thus required to extract properly the sole Overhauser shift. In that purpose, we used the spectra shown in Fig.~\ref{Fig1}(a) which are assumed to be DNP free, since they were performed under non-resonant and non-polarized excitation.

As an illustration of this procedure, Fig.~\ref{Fig4}(b) shows the raw  splittings of QD1 $\sigma^+$ and $\sigma^-$ lines for different measurements and after subtracting  the  linear slope due to the  \xp Zeeman effect ($\approx$170 \textmu eV/T).  The  $|\pm1\rangle$  anti-crossings give rise to  the same  specific  profile in all measurements which enables us to extract  the superimposed  Overhauser shift created under circularly polarized excitation  as reported  in Fig.~\ref{Fig4}(c). Note that this is the Overhauser shift of the \xp transition which, besides the dominant shift of the electron Zeeman splitting, can also  include a contribution from the hyperfine interaction with the QD-confined hole\cite{Chekhovich2011-PRL,Chekhovich2011-NatPhys,Fallahi2010}. To asses its relative size, we can compare the energy jumps  of the \xp bright transitions ($|\uparrow \Downarrow \Uparrow\rangle\rightarrow|\Uparrow\rangle$) with the  ``dark" transitions ($|\uparrow \Downarrow \Uparrow\rangle\rightarrow|\Downarrow\rangle$). The latter are partially visible thanks to the \a0-induced coupling of the hole spins\cite{Kudelski2007,Krebs2009,Goryca2010}. For QD1, we found that the energy jump  of the dark transition marked in Fig.~\ref{Fig4}(a) is reduced by about 10\% with respect to the jumps of the bright transitions, from which we deduced a $\approx$5\%  positive contribution of the hole hyperfine interaction  to the measured Overhauser shift.

%Jusque l\`{a}, le OHS du trou n'\'{e}tait pas indispensable, mais l\`{a} il faut commencer \`{a} prendre position...

At the first glance, the build up of the Overhauser shift in Fig.~\ref{Fig4}(c) turns out  similar to that observed in undoped QDs, with a roughly linear increase in positive fields up to a few Teslas. However,  the Overhauser shift lies significantly above the  electron Zeeman splitting $-g_e^\star \mu_\text{B} B_z$ (dotted line in Fig.~\ref{Fig4}(c)), where $g_e^\star=1.05 g_e$ is used to take into account the 5\% hole contribution. In undoped QDs, the distance to the Zeeman splitting is usually less than $\simeq$10 \textmu eV, in order to satisfy the self-consistent condition enabling a high DNP rate over the build up range \cite{Urbaszek2013}. Obviously, the relevant electron spin splitting must include   the electron-\a0 exchange which, for the $|\!-1\rangle$ levels, corresponds  to a shift by   $B_\Delta=\Delta_{e-A^0}/g_e \mu_\text{B}\approx-1$~T of the Zeeman splitting. The agreement in Fig.~\ref{Fig4}(c) is indeed better with the  dashed line $-g_e^\star \mu_\text{B}(B_z +B_\Delta)$, or  with the theoretical electron spin splitting (solid line) deduced from the $|\!-1,\uparrow \rm{or} \downarrow\rangle$ levels of Fig.~\ref{Fig1}(b) which also includes the $\delta$ coupling near zero field. %, yet recalculated with $g_e^\star$ instead of $g_e$.

The Overhauser shift remains however   below the electron spin splitting by  $\simeq 10$~\textmu eV, while  in undoped QDs it usually exceeds  the Zeeman splitting by about the same amount.  We believe this is due to the electron spin depolarization by the $\delta_e/2$  effective coupling  which slightly changes the stability point of the DNP. Indeed,  the Overhauser shift cannot grow above the total spin splitting  (including the Zeeman and exchange terms) because it would  require to go through the $\delta_e$ anti-crossing, where the electron spin projection $S_{e,z}$ vanishes. In other words, when approaching the anti-crossing splitting $\delta_e$,  the DNP rate is drastically reduced, in such a way that  the maximum Overhauser shift remains  slightly below the electron spin splitting by a few  $\delta_e$.

Conversely, the $\delta_e$-induced depolarization observed in Fig.~\ref{Fig2} is significantly changed due to the Overhauser shift. For example, the cross-polarized spectra in Fig.~\ref{Fig4}(a)  no longer exhibit any enhancement around +1.15~T. The $|\!+1,\uparrow\rangle$ and $|\!+1,\downarrow\rangle$ levels are indeed significantly split by the additional nuclear field in this region and the corresponding $\sigma^-$ and $\sigma^+$ \xp lines  keep a high $\approx$9:1 intensity ratio, whereas they have basically the same intensity in Fig.~\ref{Fig2}(d). More generally,   the  relative changes in intensity of the lines in Fig.~\ref{Fig4}(a) seem well correlated  to  the changes of  the measured Overhauser shift which determine the precise splitting of the electron spin states and therefore their actual mixing.
% En particulier on peut tracer l'origine des fluctuations de polar de la raie -1 au splitting total de l'\'{e}lectron, vers 200mT et vers la fin du champ Overhauser.
%Faire un nouveau paragraphe en partant de la sym\'{e}trie des branches +1 et -1 pour poser la question si une DNP d\'{e}cal\'{e}e en champ positif associ\'{e}e \`{a} la branche -1 ne serait pas possible. L\`{a} on discute un possible effet observ\'{e} sur QD1  d'encha\^{\i}ner sur QD3 (faible \'{e}change) et QD4 (fort \'{e}change mais faible \delta_e).
% l'approche symm\'{e}trie ne convient pas car elle tend \`{a} imaginer un renversement de tous les spins et du champ.

\section{Replica of DNP sequence due to \a0-electron exchange}

So far, we discussed the Overhauser field developing due to electron-nucleus flip-flops taking place between the $|\!-1,\uparrow \rm{or} \downarrow\rangle$ levels. There is an obvious question whether the DNP could  take place  with the $|\!+1,\uparrow \rm{or} \downarrow\rangle$ levels under the same $\sigma^-$ polarized excitation.
%which determines the sign of the Overhauser field and subsequently the positive field range where the DNP can develop.
The main  difference between the two configurations is that the $|\!+1,\uparrow \rm{or} \downarrow\rangle$ levels become significantly less populated than the  $|\!-1,\uparrow \rm{or} \downarrow\rangle$ ones when a positive field increases because of \a0 thermalization (see Fig.\ref{Fig1}(d)). The DNP is thus expected to be less efficient for \a0 in $|\!+1\rangle$ state. In  most of the investigated QDs there is indeed no evidence of such contribution.  By analyzing the small Overhauser shift of QD1 under  excitation at E$_1$ energy, we  found  possible indications of two  distinct increases of DNP field, but with a rather poor signal to noise ratio. Clear DNP replica were   however   observed for two  other quantum dots, QD3 and QD4, as discussed below.

\begin{table}[h!]% add [h] placement to break table across pages
\caption{\label{param}Parameters extracted from measurement analysis. Units are mT for $|B_{\delta}|$ and \textmu eV for $\Delta_{\text{e-A}^0}$ and $\delta_e$.  }
\label{ParamTable}
\begin{ruledtabular}
\begin{tabular}{c | c c c c c c }
QD$\sharp$ & $g_{A^0}^\text{eff.}$ & $|B_{\delta}|$ & $\Delta_{\text{e-A}^0}$ & $g_{e}$  & $g_{e}^\star$  & $\delta_e$\\ %  &  OHS$_\text{hole}$ le gg_A0^eff inclut le \cos \theta_s du model. C'est le facteur g exp\'{e}rimental d\'{e}duit des croix.
%\hline
\hline
QD1 &   3.6 & 79 & 33 & -0.48  & -0.504   & 9\\  %QD49 & +5\%
QD2  &   2.8 & 215 & 72 & -0.57 &  -  & 47 \\ %QD287 & -
QD3 &   3.7 & 22 & 10 & -0.52  & -0.556  & 4 \\%QD113 & +7\%
QD4 &   3.4 & 50 & 20 & -0.51  & -0.42  & 8 \\% QD368  & -17\% Le B_delta extrait de sp571 est plut\^{o}t 42mT, mais je fais plus confiance aux mesures bas\'{e}es sur le splitting \'{e}lectronique en fort champ magn\'{e}tique, et \c{c}a marche mieux pour fitter le OHS avec un B_ac de 0.7T\approx 50 2 3.4/0.51.
\hline
 \end{tabular}
 \end{ruledtabular}
\end{table}

\begin{figure}[ht]
\includegraphics[width=0.48\textwidth]{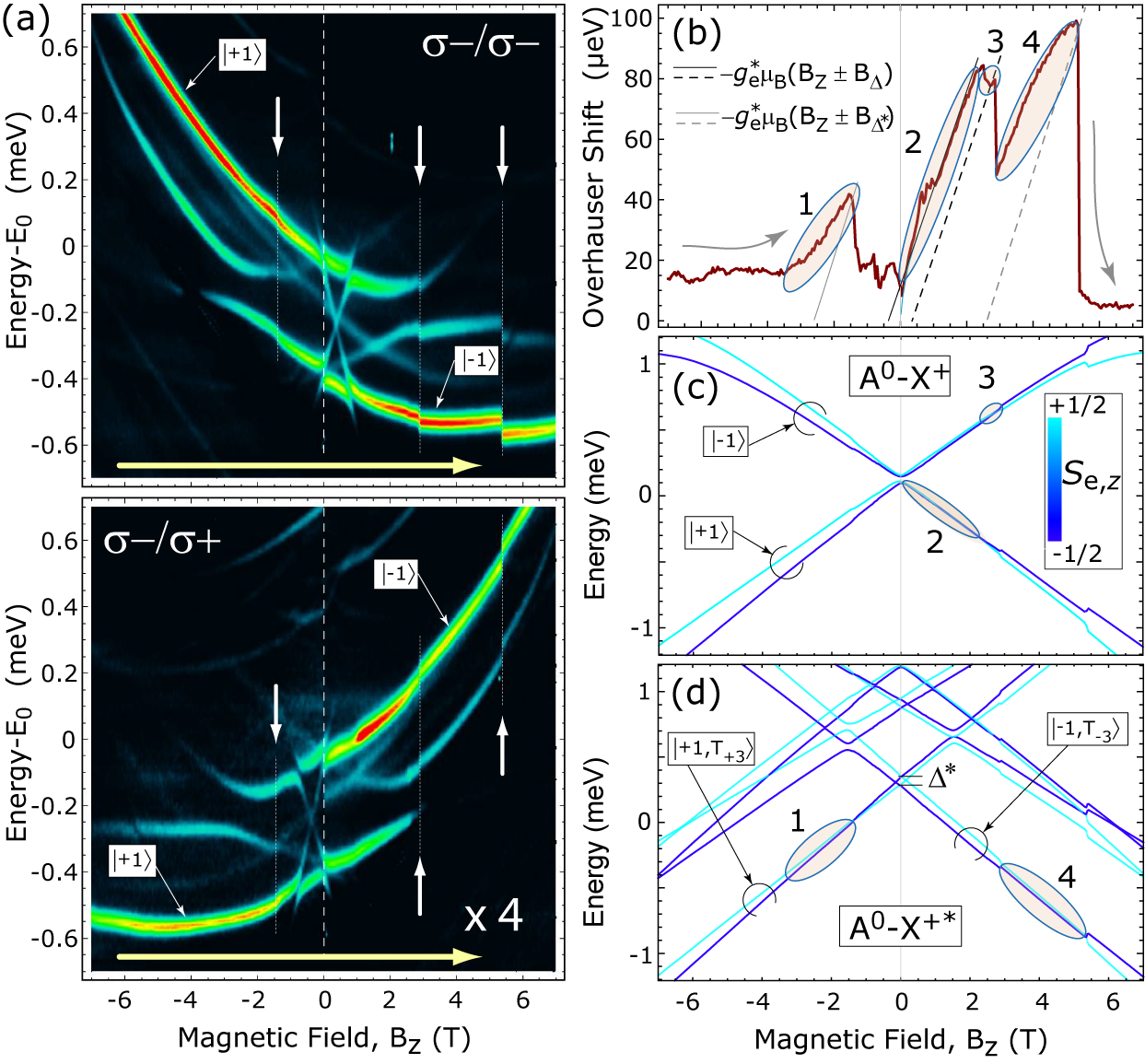}
\caption{\label{Fig5} (a) QD3 PL spectra under quasi-resonant ($E_0$+72~meV)  $\sigma-$ excitation   as a function of an increasing  magnetic field $B_z$   and measured in $\sigma-$ (top) or $\sigma+$ (bottom) polarization ($E_0=1.304$~eV). (b) Overhauser shift deduced from (a) and showing four distinct DNP ranges. (c), (d) Calculated energy levels in the states A$^0$-X$^+$  (c)  and A$^0$-X$^{+^\star}$ (d) including the experimental Overhauser shift. Each DNP sequence in (b) can be associated to a pair of  closely spaced electron spin levels .
}
\end{figure}
\begin{figure}[ht]
\includegraphics[width=0.48\textwidth]{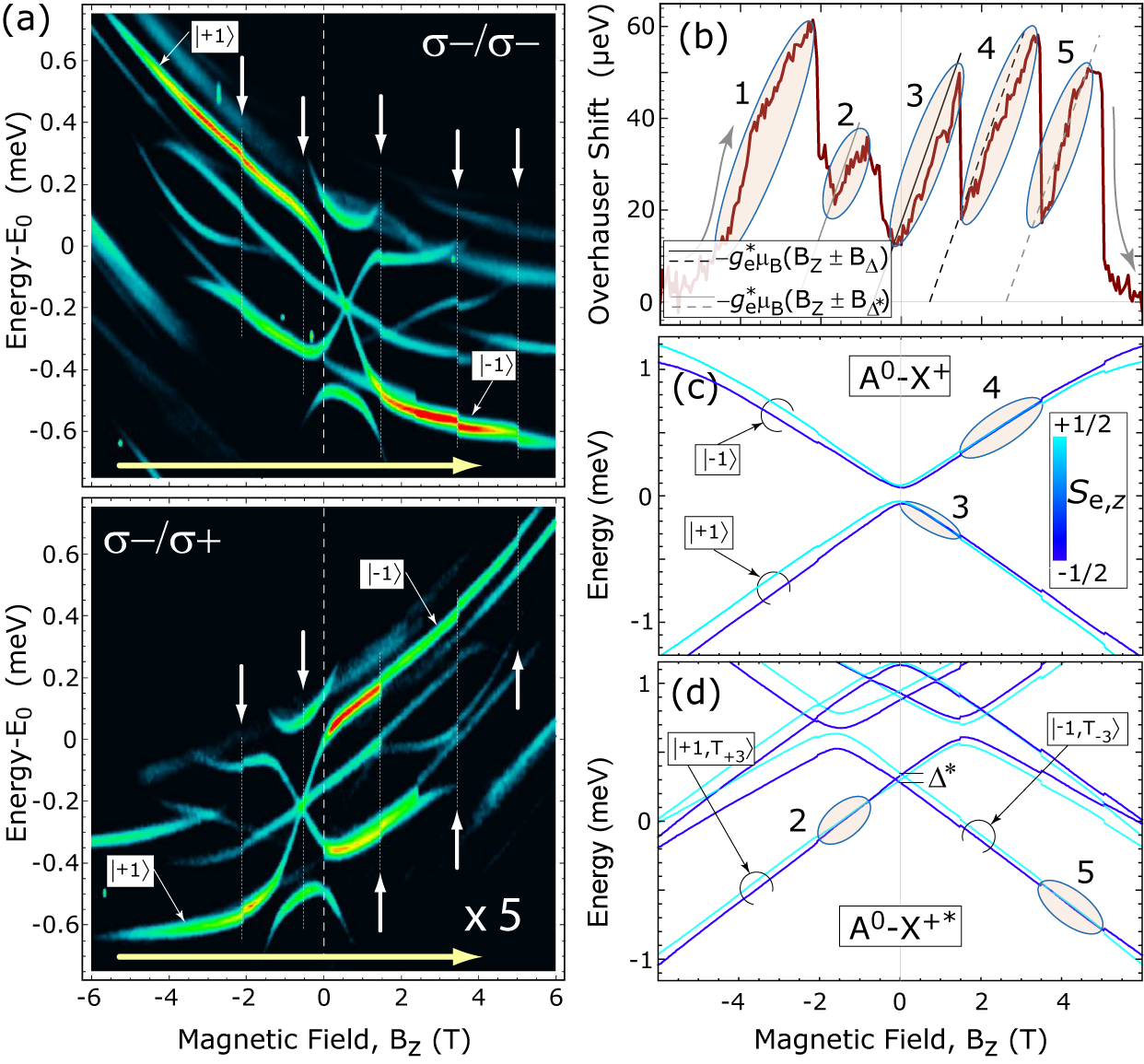}
\caption{\label{Fig6}  (a) QD4 PL spectra under quasi-resonant  ($E_0$+76~meV) $\sigma-$ excitation   as a function of an increasing  magnetic field $B_z$   and measured in $\sigma-$ (top) or $\sigma+$ (bottom) polarization ($E_0=1.307$~eV). (b) Overhauser shift deduced from (a) and showing five successive DNP sequences. (c), (d) Calculated energy levels in the state A$^0$-X$^+$  and A$^0$-X$^{+^\star}$ including the experimental Overhauser shift. The DNP sequences in (b)  can be associated to a pair of  closely spaced electron spin levels, but the first increase `1' which would require another trion state.
}
\end{figure}

The DNP measurements and analysis of QD3 and QD4 are respectively shown in Fig.~\ref{Fig5} and Fig.~\ref{Fig6}. Both QDs were quasi-resonantly excited at about 2 GaAs LO-phonon energy ($\approx$74 meV) above the ground state emission at $\approx$1.3~eV.~\footnote{A lower excitation at about 1 LO-phonon, like for QD1, gave the same results for QD3, but  was unable to create efficiently the \xp state in QD4.} By sweeping the magnetic field from -6~T to +6~T under a constant $\sigma^-$ polarized excitation, we observed several spectral jumps of the QD PL lines, to lower or higher energy depending  on the co- or cross-polarized detection configuration, indicating obviously several DNP sequences. By following the same procedure as for QD1, we extracted a quantitative estimate of the Overhauser shift with respect to a DNP-free reference spectrum (not shown).  Remarkably, the  figures~\ref{Fig5}(b) and~\ref{Fig6}(b) show that both QDs experience successive increases of nuclear polarization, each  interrupted by an abrupt,  partial or total fall   and  exhibiting roughly the same slope.

Observing more than two increases was really not anticipated. Indeed, this can not be interpreted in the frame of a DNP rate determined by the electron spin-splitting of only the \xp-\a0 ground  levels. We believe that other levels are required and   suggest tentatively in the following that hot trion states X$^{+^\star}$ could be responsible for the DNP increases starting in  high (positive or negative) magnetic fields due to an additional electron-hole exchange energy\cite{Warming2009a,Siebert2009}. First, let us focus  on the DNP increases starting from a smaller field $|\pm B_\Delta|<$2~T, that we ascribe to the \xp-\a0 levels.

% dire petit \delta_e, refaire dessin pour contrib niveau +1.
%The main differences are a lower e-\a0 exchange for QD3 and an inversion of the hole contribution to the Overhauser shift for QD4, with respect to QD3 and QD1.

The theoretical \xp-\a0  levels are  plotted in Fig.~\ref{Fig5}(c) and Fig.~\ref{Fig6}(c) from  a spin model using the parameters deduced from the experiments, but including also  the measured Overhauser shift. This points out different parts of the $|\pm1\rangle$ levels (shaded area in the figures) where the actual electron spin splitting is particularly small and therefore enables fast electron-nuclei flip-flops. The relevant g-factors and exchange energy of QD3 and QD4,  deduced from the anti-crossing fields $B_\delta$ or the splitting between bright and dark lines are put together in Tab.~\ref{param}.

For QD3, the electron-\a0 exchange is small (10 \textmu eV) and thus produces only a small difference of  electron spin splitting between the $|\!+1\rangle$ and $|\!-1\rangle$ \a0 states. The DNP associated to the  $|\!-1,\uparrow \rm{or} \downarrow\rangle$ levels [labeled `2' in \ref{Fig5}(b)] actually exhibits two regimes : the Overhauser shift first exceeds the electron spin splitting $\approx g_e^\star \mu_\text{B} (B_z+B_\Delta)$ in a way similar to undoped QDs, then they cross each other around 1~T but remain very near in energy up to 2.1~T. This is confirmed by the sudden increase at 1~T  of the $|\!-1\rangle$ line PL intensity in cross-polarized configuration in Fig.~\ref{Fig5}(a) due to the $\delta_e$ induced coupling. Even though this regime is similar to that observed for QD1, the DNP mechanism likely  benefits here from the  proximity (less than 20 \textmu eV) of the $|\!+1,\uparrow \rm{or} \downarrow\rangle$ spin splitting  $\approx g_e^\star \mu_\text{B} (B_z-B_\Delta)$. Indeed, above 2.1~T the Overhauser shift starts a decrease soon interrupted by the DNP sequence `3' associated to the $|\!+1\rangle$ levels. It remains at a rather high level by roughly following the  corresponding spin splitting   up to 2.9~T, where only  it experiences a large reduction by about 30~\textmu eV.

For QD4, the two DNP increases associated to the two \a0 states (labeled `3' and `4' in Fig.~\ref{Fig6}(b)) are even more clearly identified  thanks to a larger electron-\a0 exchange  (20 \textmu eV) which determines their starting fields $\pm B_\Delta$ at $\approx \pm 0.7$~T. It is noteworthy that the effective g-factor $g_e^\star$ used to reproduce their slope is smaller (in absolute value) than the actual electron g-factor, see Tab.~\ref{param}. Like for QD1 and QD3, it was determined from  the estimate of the hole contribution to the Overhauser shift   by comparing the spectral jump  of a dark line with respect to a bright line. For   QD4 this contribution was surprisingly found to be negative (by about -17\%) in contrast to QD1 and QD3. This effect likely results from a very different Indium composition of QD4 combined with the different hyperfine coupling constants  of the host atomic species, or possibly from a very different strain profile affecting drastically these constants~\cite{Chekhovich2011-NatPhys}. Further investigations are required to answer this issue and Mn-doped QDs could reveal particularly interesting  in that respect by providing almost systematically measurable dark transitions.

\begin{figure}[ht]
\includegraphics[width=0.48\textwidth]{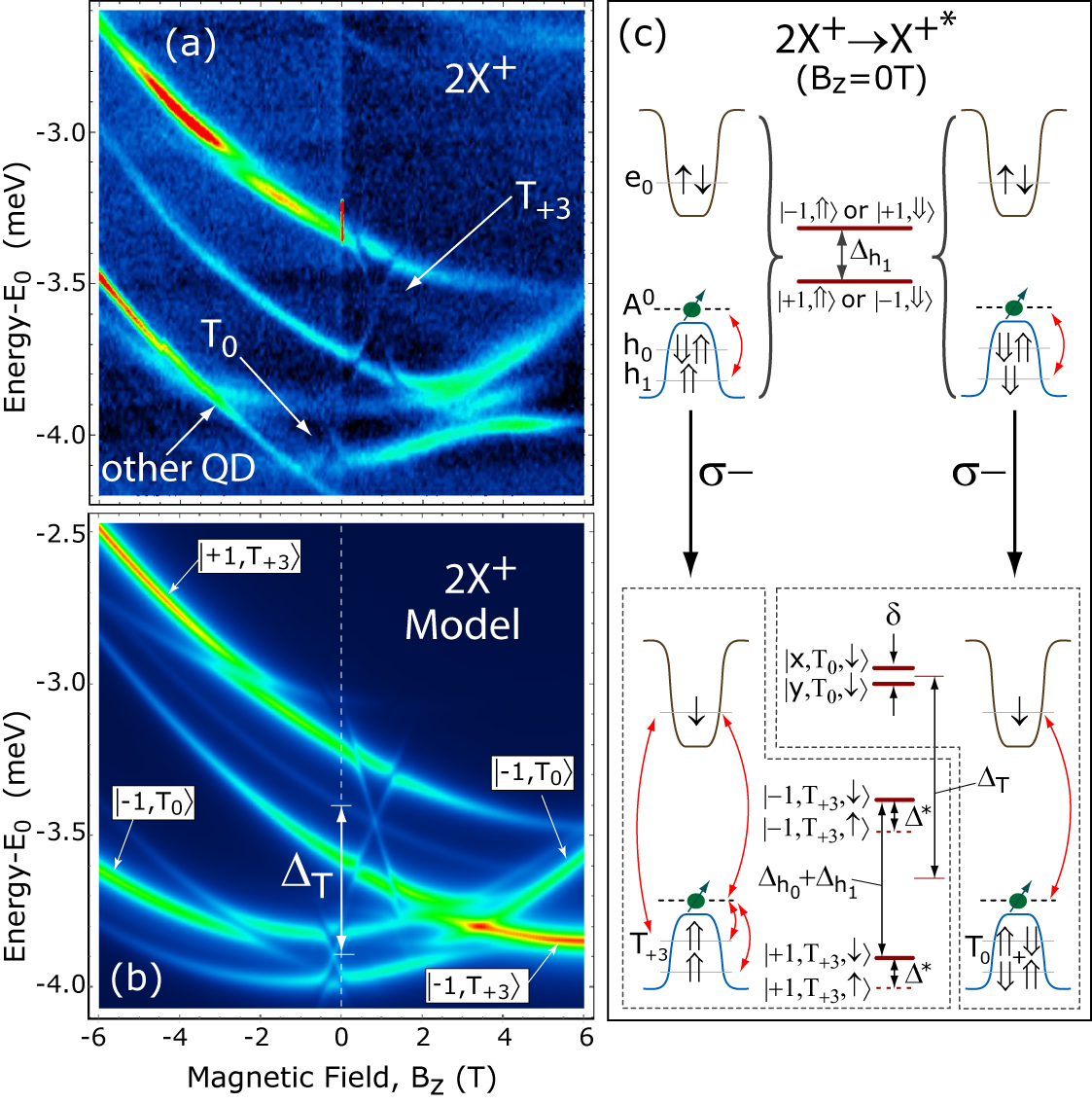}
\caption{\label{Fig7}  (a) QD3 PL spectra under quasi-resonant ($E_0$+72~meV)  unpolarized excitation   as a function of the magnetic field $B_z$   and measured in $\sigma^-$ circular polarization ($E_0=1.304$~eV). (b) Simulated magneto-PL spectra of a \2xp$ \rightarrow$\xps  transition, from a standard spin  model. The main fitting parameters are the triplet $T_0$-$T_{\pm3}$ splitting ($\Delta_\text{T}$=0.5~meV), the \a0 exchange energies with  the ground (excited) hole state $\Delta_{h_{0(1)}}=0.4$ (0.15)~meV  and  the hole g-factors $g_{h_{0(1)}}=2.25(-1.05)$. (c) Schematics of the spin levels and exchange interactions (red arrows) involved in a \2xp$ \rightarrow$\xps  $\sigma^-$-polarized transition at zero field.
}
\end{figure}

\section{tentative interpretation of DNP replica in high fields}

The increases occurring in higher fields and  denoted `1', `4' in Fig.~\ref{Fig5}(b) and `1', `2', `5' in Fig.~\ref{Fig6}(b)  remain truly surprising because they cannot be understood from the level structure in  the \a0-\xp configuration. Their slopes, close to $g_e^\star\mu_\text{B}$, still suggest a  DNP mechanism based on the compensation by the Overhauser field of an electron spin-splitting that would be shifted to higher (positive or negative) fields due to the exchange with a  different configuration of the other present spins. Indeed, for both QD3 and QD4 we can trace back a symmetrical origin for  two of these DNP increases at the starting field $\pm B_{\Delta^\star}\approx \pm2.5$~T   which can be ascribed to an exchange energy  $\Delta^\star= g_e \mu_\text{B} B_{\Delta^\star}\approx 75$~\textmu eV [see the gray lines in Fig.~\ref{Fig5}(b) and  Fig.~\ref{Fig6}(b)].

To support the interpretation of the high field DNP replica, we considered as a probable candidate an \a0-\xps state, where \xps is a hot \xp trion having one of its two QD-confined holes occupying a QD excited level. It can be  created under non-resonant PL excitation of undoped InGaAs QDs in particular as the intermediate state of a charged biexciton (\2xp) cascade\cite{Warming2009a,Siebert2009,Poem2010a}. For QD3, we observed a group of spectrally correlated lines, about 3.5~meV below the dominant \a0-\xp features, that we indeed identified as a \2xp$ \rightarrow$\xps transition, thanks to a theoretical simulation  of the  magneto-PL image, see Fig.~\ref{Fig7}. It basically consists of two pairs of $|\pm1\rangle$ lines which can be associated to the $T_{+3}=|\Uparrow\Uparrow\rangle$ or $T_{0}=(|\Uparrow\Downarrow\rangle+|\Downarrow\Uparrow\rangle)/\sqrt2$  triplet configurations of a ground state hole ($h_0$) and an excited state hole ($h_1$) split by the  hole-hole exchange $\Delta_\text{T}$\cite{Ediger2007,Warming2009a}. They each exhibit a specific X-pattern which differs in field position and size. In  $T_{+3}$ configuration, the excited $h_1$ hole has the same spin in the initial and final state so that its exchange $\Delta_{h_1}$ with \a0 essentially produces a shift in magnetic field by $\Delta_{h_1}/g_{A^0} \mu_\text{B}$  of an X-pattern similar to the \xp one. In  $T_{0}$ configuration, both holes have zero spin projection along $z$ so that their exchange  with \a0 vanishes. The corresponding X-pattern is essentially determined by the $\Delta_{h_1}$  exchange  in the initial state, see  Fig.~\ref{Fig7}(c). Note that under a fixed polarization  detection ($\sigma^-$ here) there is a correlation  between the $h_1$ $\Uparrow$ or $\Downarrow$ spin  and the $T_{+3}$ or $T_0$  triplet state\cite{Poem2010a}, which explains the position in positive or negative field of the corresponding X-pattern. Finally,  the electron-hole exchange which also contributes to the triplet splitting $\Delta_\text{T}$, determines with the electron-\a0 exchange, the electron spin splitting $\Delta^\star$  between the bright ($|\pm1,T_{+3},\downarrow\rangle$) and   dark  ($|\pm1,T_{+3},\uparrow\rangle$)  configurations of \xps as represented in Fig.~\ref{Fig7}(c).

Theoretically, this $\Delta^\star$ exchange splitting could control the  electron-nucleus flip-flop rate in the \xps state and thus explain the additional  increases starting from the fields $\pm \Delta^\star/\ g_e \mu_\text{B}$. However, in contrast to Ref.~\onlinecite{Warming2009a}, we could not identify  any spectral lines associated to  the \xps dark states, and therefore  the value of  $\Delta^\star$ could not be determined  experimentally. Therefore, to complete our tentative interpretation we  simply adjusted  $\Delta^\star$  in order to keep the actual electron spin splitting (including the measured Overhauser shift) as small as possible  over the field ranges where the increases develop. The corresponding calculated levels are shown in Fig.~\ref{Fig5}(d) and~\ref{Fig6}(d) for QD3 and QD4, with shaded areas emphasizing the $|\pm1,T_{\pm3},\uparrow \text{or} \downarrow\rangle$ levels where the spin splitting is  less than $\approx$25~\textmu eV.

The value of  $\Delta^\star\approx$75~\textmu eV determined in this way  turns to be about one order of magnitude smaller than the value reported in Ref.~\onlinecite{Warming2009a} for InAs/GaAs QDs. If such discrepancy is confirmed in future investigations, this might obviously question our specific interpretation  based  on an \xps state, but the principle  should remain valid for another excited state to  be identified. This is besides the case of the first DNP increase of QD4   (labeled `1' in Fig.~\ref{Fig6}(b)) which starts at $\approx$-5~T  and therefore cannot be explained by any of the spin splittings calculated for \xp or \xps. Since so far no such effects have been observed  with undoped QDs, there is still a strong suspicion that the magnetic impurity is  solely responsible for all of the observed DNP increases. This could be due to  variations of the electron-\a0 exchange energy either for an excited  electron occupying a different QD orbital, or for an excited  \a0  spin configuration  (notably in a $J=2,3 \text{ or } 4$ spin state).
%Mentionner ici que les transitions J=2 sont observ\'{e}es pour QD49 et QD287 ?

\section{summary}

In conclusion, our investigations of the optical orientation of \xp trions in Mn-doped InGaAs QDs and the subsequent dynamic nuclear polarization  have revealed the key role played by the exchange interaction $\Delta_e$ between the QD-confined electron and  the Mn-induced neutral acceptor state \a0. On the one hand, the anisotropic part of this exchange gives rise to an effective direct coupling of the electron spin states resulting in the anticrossing $\delta_e$ of the corresponding levels and the vanishing of the \xp spin orientation at the specific magnetic fields $\pm\Delta_e/ g_e \mu_\text{B}$. This  limits  the maximum  of the Overhauser field  due to the reduction of the electron nuclei flip-flop rate when the  Overhauser shift approaches the  electron Zeeman splitting. On the other hand, the longitudinal (Ising-like) part of the exchange acts as an effective magnetic field along $z$ giving rise to two successive DNP increases. This analysis is supported by a precise determination of the electron g-factor and the contribution of the hole spin to the Overhauser shift,   evidencing an  energy-driven  DNP mechanism similar to that observed for undoped QDs. More surprising is the observation for certain QDs of additional increases in higher  positive or  negative magnetic fields. We propose an interpretation based on a  DNP mechanism taking place in an \xps state and shifted to a higher  fields due to the exchange interaction between the electron and two holes in a triplet $T_{\pm3}$ configuration. Still, alternative explanations involving  the \a0 excited states remain plausible and should be more specifically investigated in future works.

%Dire aussi que le X+* peut \^{e}tre form\'{e} directement.  Il reste difficile de justifier que l'on forme autant de X+* en config T+3 que T-3...
%s'appuyer sur les papiers de Bimberg pour dire que les deux \'{e}tats triplets 7/2 et 5/2 sont proches en \'{e}nergie.
% mentionner que le facteur g du trou P est n\'{e}gatif et plus faible. mais ces d\'{e}tails sont out of the scope of the present paper.

\begin{acknowledgments}
This work was partially supported by ``Triangle de la Physique'' (project COMAQ), the French RENATECH network and the ``Labex NanoSaclay'' (project ICQOQS).
\end{acknowledgments}

\bibliography{biblioQDMn}

\end{document}